# MENP: An Open-Source MATLAB Implementation of Multipole Expansion for Applications in Nanophotonics

Tatsuki Hinamoto* and Minoru Fujii*

Department of Electrical and Electronic Engineering, Graduate School of Engineering,
Kobe University, Kobe 657-8501, Japan
*tatsuki.hinamoto@gmail.com
*fujii@eedept.kobe-u.ac.jp

**ABSTRACT**

In modern nanophotonics, multipolar interference plays an indispensable role to realize novel optical devices represented by metasurfaces with unprecedented functionalities. Not only to engineer sub-wavelength structures that constitute such devices but also to realize and interpret unnatural phenomena in nanophotonics, a program that efficiently carries out multipole expansion is highly demanded. MENP is a MATLAB program for computation of multipole contributions to light scattering from current density distributions induced in nanophotonic resonators. The main purpose of MENP is to carry out post-processing of a rigid multipole expansion for full-field simulations which in principle provide the information of all near- and far-field interactions (*e.g.* as a total scattering cross section). MENP decomposes total scattering cross sections into partial ones due to electric and magnetic dipoles and higher-order terms based on recently developed exact multipole expansion formulas. We validate the program by comparing results for ideal and realistic nanospheres with those obtained with the Mie theory. We also demonstrate the potential of MENP for analysis of anapole states by calculating the multipole expansion under the long-wavelength approximation which enables us to introduce toroidal dipole moments.

Nature of problem: To quantitatively and qualitatively clarify the origin of scattering behaviors accompanied by electromagnetic resonances of sub-wavelength structures.

Solution method: The multipole expansion of induced current density distributions based on an exact expression recently derived. Besides, an expression under the long-wavelength approximation is also utilized to obtain multipole moments including a toroidal dipole.

Additional comments including Restrictions and Unusual features: MENP has been developed primarily for combination with FDTD solutions (Lumerical inc.). A Lumerical script file which provides packaging of the needed data, including electric field and refractive index distributions, into MATLAB is included in the MENP package. Nonetheless, any simulation software can be used to produce electric field distributions by exporting them as a MATLAB (.mat) file.

1. **INTRODUCTION**

The emergence of all-dielectric nanophotonics as an alternative to plasmonics has opened up a door toward the engineering of scattering behaviors by sub-wavelength nanophotonic resonators with an unprecedented degree of freedom.[1–5] The operation of all-dielectric nanophotonics relies on Mie resonances accompanied by light confinement within sub-wavelength nanostructures made of high-refractive-index materials. One of the important features of the Mie resonance in stark contrast to plasmonics is the possession of electric and magnetic dipole and higher-order multipole resonances.[6–9] It has been proposed and demonstrated that engineering of these modes can remarkably improve the performance of optical components including metasurfaces with the help of the low-loss nature of all-dielectric Mie resonators.[10] The development of all-dielectric nanophotonics could impact a wide range of fields, including optical elements,[11] detectors,[12,13] light-sources,[14,15] sensing[16], nanophotonic inks,[17] and so on.

It has been demonstrated that multipolar interference between a series of resonant modes leads to fascinating phenomena.[10] For example, the Huygens source, an individual scatterer which radiates light only into forward direction, can be realized with non-magnetic structures by the far-field interference between electric and magnetic dipole scatterings.[18–22] Such zero-backward-scattering occurs at the condition where the electric ($p$) and magnetic ($m$) dipole moments have the same scattering amplitudes with in-phase (so-called the first Kerker condition: $p - m/c = 0$, where $c$ is the speed of light.).[21] The concept of the Kerker-type directionality has been extended into almost zero-forward scattering (the anti-Kerker condition: $p + m/c = 0$) and transverse scattering.[23] Another intriguing phenomenon is an anapole state which exhibits no radiating field; the phenomenon is interpreted as a result of destructive interference between electric and toroidal dipole moments.[24] Note that several different interpretations are possible for the anapole state.[25–28] The radiationless state is of importance in realizing invisible photodetectors with minimum cross-talk between the other elements and to confine electromagnetic energy inside subwavelength resonators.[28–33]

Because most of the fascinating phenomena in nanophotonics, including the above mentioned Kerker condition and non-radiating anapole state, stem from multipolar interferences of scattered fields between multiple optical resonances, an analysis based on multipole expansion is essential for understanding and engineering optical properties of nanophotonic resonators.[34–36] Although analytical electromagnetic calculation such as the Mie theory[37] naturally includes the multipole analysis, no analytical expressions of optical responses are available for most structures except for some simple ones such as a sphere. For the computation of electromagnetic responses of complex structures, full-field simulation techniques based on finite-difference time-domain (FDTD) method, finite-element method (FEM), boundary element method (BEM), discrete-dipole approximation (DDA), and so on are thus utilized. In commercially available simulation software, it is straightforward to obtain electric and magnetic field distributions and some typical quantities such as *total* scattering and absorption cross sections. However, decomposition of the total scattering into multipole contributions can be problematic because such analysis is not often provided in public, requiring us more or less effort for coding of complex expressions. To this end, a program that efficiently computes multipole expansion from the electric field distributions is highly demanded.

In addition to the well-known formulation of multipole expansion found in textbooks of electrodynamics,[38] some expressions have been developed for easier implementation in designing optical resonators.[27,34–36] The formulations can be classified into several methods depending on the basis (Spherical or Cartesian) and the approaches (scattered fields or induced currents), and those based on induced current density distributions in the Cartesian coordinate system is often adopted in nanophotonics applications. Up to now, an exact expression recently derived by Alaee and coauthors is, in our opinion, one of the most straightforward formulations.[35] Besides, the expression derived under the long-wavelength approximation is also commonly utilized with the introduction of *so-called* toroidal moments which are higher-order terms of electric moments.[27,39]

Inspired by the work by Alaee, *et al*. and substantial demands in the field, here we develop an open-source MATLAB code that computes multipole expansion for nanophotonics (MENP). MENP provides an efficient computation using matrix processing in MATLAB to treat four-dimensional (4D) matrices of electric field distributions. This work consists of the following sections. After summarizing the theoretical expressions implemented in MENP, we explain the overview of the program. We then validate our implementation by computing the exact multipole expansion for a lossless nanosphere with a combination of FDTD and MENP, followed by a comparison to analytically obtained results. The procedure is also carried out for a silicon nanosphere with complex refractive indices. Finally, we apply MENP to a silicon nanodisk to show the existence of the anapole state based on the multipole expansion under the long-wavelength approximation. Although the MENP has the best compatibility with FDTD solutions (Lumerical Inc.) because of the authors' simulation environment, it is a versatile program and is not limited to the combination with FDTD. MENP will significantly contribute to nanophotonics fields by providing physical insights into multipolar interferences in various structures.

2. **THEORY**

First, we describe the exact expression of multipole expansion reported in [35]. As a starting point, let us define a problem: a resonator in free space is illuminated by a plane wave with an electric field amplitude $|\boldsymbol{E}_{\text{inc}}| = E_0$ at the frequency $f$. The basis is a Cartesian coordinate system, and a position vector can be defined as $\boldsymbol{r} = (x, y, z)$. When incident light excites the resonator, the induced current density distributions $\boldsymbol{J}(\boldsymbol{r})$ can be obtained from the electric field distributions $\boldsymbol{E}(\boldsymbol{r})$ by

$$\boldsymbol{J}(\boldsymbol{r}) = -i\omega\varepsilon_0(n^2 - 1)\boldsymbol{E}(\boldsymbol{r}) \tag{1}$$

, where $\omega$ is the angular frequency, $\varepsilon_0$ is the permittivity of free space, and $n$ is the refractive indices of the resonator. Note that $\boldsymbol{J}(\boldsymbol{r})$ corresponds to displacement current distributions in the case of dielectric Mie resonators. The multipole moments, that is, electric dipole ($\boldsymbol{p}$), magnetic dipole ($\boldsymbol{m}$), electric quadrupole ($\hat{Q}^{\text{e}}$), magnetic quadrupole ($\hat{Q}^{\text{m}}$), can be derived as [35]:

$$\begin{aligned} p_\alpha &= -\frac{1}{i\omega}\left[\int J_\alpha j_0(kr)\, d^3\boldsymbol{r} + \frac{k^2}{2}\int \{3(\boldsymbol{r}\cdot\boldsymbol{J})r_\alpha - r^2 J_\alpha\}\frac{j_2(kr)}{(kr)^2}\, d^3\boldsymbol{r}\right] \\ m_\alpha &= \frac{3}{2}\int (\boldsymbol{r}\times\boldsymbol{J})_\alpha \frac{j_1(kr)}{kr}\, d^3\boldsymbol{r} \\ \hat{Q}^{\text{e}}_{\alpha\beta} &= -\frac{3}{i\omega}\bigg[\int \{3(r_\beta J_\alpha + r_\alpha J_\beta) - 2(\boldsymbol{r}\cdot\boldsymbol{J})\delta_{\alpha\beta}\}\frac{j_1(kr)}{kr}\, d^3\boldsymbol{r} \\ &\quad + 2k^2\int \{5r_\alpha r_\beta(\boldsymbol{r}\cdot\boldsymbol{J}) - r^2(r_\alpha J_\beta + r_\beta J_\alpha) \\ &\quad - r^2(\boldsymbol{r}\cdot\boldsymbol{J})\delta_{\alpha\beta}\}\frac{j_3(kr)}{(kr)^3}\, d^3\boldsymbol{r}\bigg] \\ \hat{Q}^{\text{m}}_{\alpha\beta} &= 15\int \{r_\alpha(\boldsymbol{r}\times\boldsymbol{J})_\beta + r_\beta(\boldsymbol{r}\times\boldsymbol{J})_\alpha\}\frac{j_2(kr)}{(kr)^2}\, d^3\boldsymbol{r} \end{aligned} \tag{2}$$

, where $\alpha, \beta = x, y, z$ and $k$ is the wavenumber. Note that $j_n(\rho)$ denotes the spherical Bessel function defined by $j_n(\rho) = \sqrt{\pi/2\rho}\, J_{n+1/2}(\rho)$, where $J_n(\rho)$ is the Bessel function of first kind.

Using the above derived multipole moments, we can now calculate a total scattering cross section by[38]

$$C^{\text{total}}_{\text{sca}} = \frac{k^4}{6\pi\varepsilon_0^2|E_0|}\bigg[\sum\left(|\boldsymbol{p}|^2 + \left|\frac{\boldsymbol{m}}{c}\right|^2\right) + \frac{1}{120}\sum\left(|\hat{Q}^{\text{e}}|^2 + \left|\frac{k\hat{Q}^{\text{m}}}{c}\right|^2\right) + \cdots\bigg]. \tag{3}$$

As can be seen, the total scattering cross section is a simple sum of partial scattering cross sections from different multipoles (*i.e.*, $C_{sca}^{p}, C_{sca}^{m}, C_{sca}^{\hat{Q}^e}, C_{sca}^{\hat{Q}^m}$).

Next, we show the expression under the long-wavelength approximation. It is known that the approximated expression can be derived by making an approximation to the spherical Bessel functions.[27,35,39]. The multipole moments are expressed as:

$$p_\alpha \approx -\frac{1}{i\omega}\left[\int J_\alpha \, d^3r + \frac{k^2}{10}\int\{(r\cdot J)r_\alpha - 2r^2 J_\alpha\}d^3r\right]$$

$$m_\alpha \approx \frac{1}{2}\int (r\times J)_\alpha \, d^3r$$

$$\hat{Q}_{\alpha\beta}^e \approx -\frac{1}{i\omega}\left[\int\{3(r_\beta J_\alpha + r_\alpha J_\beta) - 2(r\cdot J)\delta_{\alpha\beta}\}d^3r\right.$$
$$+\frac{k^2}{14}\int\{4r_\alpha r_\beta(r\cdot J) - 5r^2(r_\alpha J_\beta + r_\beta J_\alpha)$$
$$\left.+ 2r^2(r\cdot J)\delta_{\alpha\beta}\}d^3r\right]$$

$$\hat{Q}_{\alpha\beta}^m \approx \int\{r_\alpha(r\times J)_\beta + r_\beta(r\times J)_\alpha\}d^3r.$$

(4)

The total scattering cross sections can be obtained in the same expression as eq. 3.

Finally, we introduce toroidal moments to the multipole family. Because the higher-order term of the electric dipole moment can be regarded as the toroidal dipole moment (**T**), the expression can be rewritten as

$$p_\alpha \approx -\frac{1}{i\omega}\int J_\alpha \, d^3r$$

$$T_\alpha \approx \frac{1}{10c}\int\{(r\cdot J)r_\alpha - 2r^2 J_\alpha\}d^3r.$$

(5)

The corresponding total scattering cross section is:

$$C_{sca}^{total} = \frac{k^4}{6\pi\varepsilon_0^2|E_0|}\left[\sum\left(|p+ikT|^2 + \left|\frac{m}{c}\right|^2\right)\right.$$
$$\left.+\frac{1}{120}\sum\left(|\hat{Q}^e|^2 + \left|\frac{k\hat{Q}^m}{c}\right|^2\right) + \cdots\right].$$

(6)

Note that here we introduced only toroidal dipole, and a higher-order toroidal moment is included in the electric quadrupole moment. It is also possible to treat the higher-order terms of the electric quadrupole moment as a *so-called* toroidal quadrupole moment. From eq. 6, the anapole condition can be derived as $p + ikT = 0$.

# 3. BENCHMARK

## 3.1. OVERVIEW

In Figure 1, we show the overview describing the calculation flow of MENP to obtain scattering spectra decomposed into multipoles. First, the electric field distributions in a resonator or resonators are simulated in an arbitrary program (Figure 1a). Typically, the simulation is carried out for a plane wave excitation with an amplitude of 1 V/m in air (*i.e.,* with a background refractive index of 1). The electric fields around the resonator should be recorded at each point in a discretized simulation mesh in three-dimension (3D; $x, y, z$) in the frequency domain ($f$), leading three ($E_x, E_y, E_z$) 4D matrices of the field distributions ($E(x, y, z, f)$). To obtain induced current distributions within the resonator, it is convenient to extract refractive index distributions at the same mesh ($n(x, y, z, f)$) because the data outside the resonator vanishes when $n = 1$ (see eq. 1). To pass the data together with arrays of coordinates ($x, y, z, f$) to MENP, they are saved all-in-one MATLAB .mat file named as ENxyzf.mat. For convenience, we provide a Lumerical script EField2MAT.lsf, which exports the needed data from a project file (.fsp) of FDTD Solutions, in /lumerical_script directory.

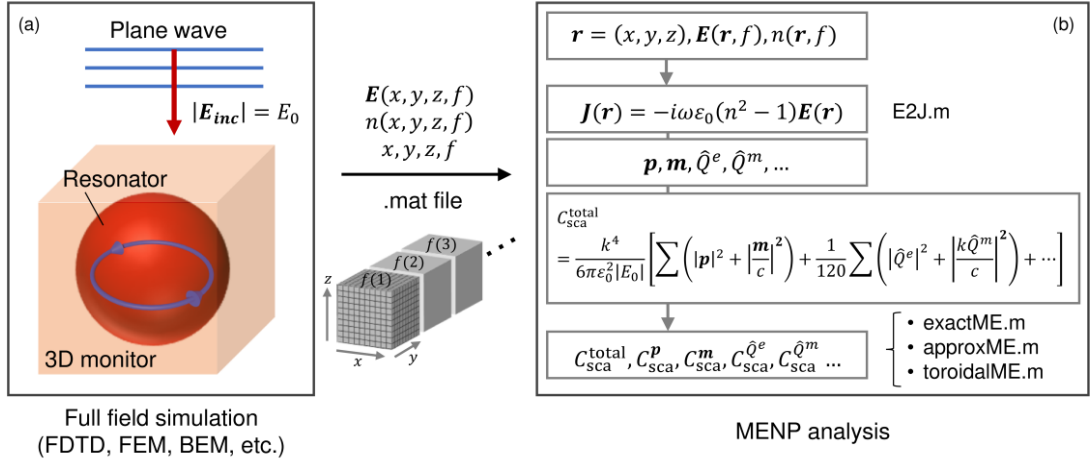

**Figure 1.** Overview of the calculation flow with MENP. (a) Calculation of electric fields ($E(x, y, z, f)$) within a nanophotonic resonator under the illumination of a plane wave ($E_{inc}$) with an amplitude $E_0$. This step is typically done using full-field simulation techniques such as FDTD, FEM, BEM, etc. (b) MENP reads the electric field distributions passed from the simulator. Note that the data should be packaged into .mat file format together with refractive index ($n(x, y, z, f)$) and one-dimensional arrays of axes ($x, y, z, f$). After the conversion of the electric field distributions into current density ones by E2J.m MENP computes multipole moments (electric dipole: $p$, magnetic dipole: $m$, electric quadrupole: $\hat{Q}^e$, magnetic quadrupole: $\hat{Q}^m$), followed by calculations of scattering cross sections.

Once the electric field distributions are obtained, MENP carries out the post-processing of multipole expansion. It is worth noting that MENP is designed for not loop-based but *vectorized* calculations of the 4D matrices, providing a better appearance of the code and faster computing in MATLAB. As shown in Figure 1b, MENP is constituted of two MATLAB functions. One is E2J.m, which converts from the electric field distributions to current density distributions based on eq. 1. The other (one of the following functions) calculates multipole moments (either exact or approximated ones) and partial and total scattering cross sections:

- exactME.m: Exact multipole expansion using eq. 2.
- approxME: Multipole expansion under the long-wavelength approximation using eq. 4.
- toridalME: Multipole expansion under the long-wavelength approximation with toroidal dipole moment using eq. 5.

They return scattering cross sections as a function of the frequency.

## 3.2. USAGE

The practical usage of MENP can be understood by looking into demo files:

/demo_sphere/demo_exact.m (based on exact expressions; eq. 2)
/demo_sphere/demo_approx.m (based on approximated expressions; eq. 4)
/demo_disk/demo_toroidal.m (based on approximated expression with toroidal dipole; eq. 5).

The installation of the MENP can be done by simply adding a path of /MENP directory to the MATLAB search path:

addpath(../MENP);

Next, one loads ENxyzf.mat in which $E, n, x, y, z, f$ are saved. The loaded variables are then passed to the main function, for example:

[Cp,Cm,CQe,CQm,Csum] = exactME(x,y,z,f,Ex,Ey,Ez,n_x,n_y,n_z);

The returned variables are partial scattering cross sections and the sum of them (*i.e.*, total scattering cross section up to quadrupoles).

## 3.3. DEMONSTRATION

We validate the MENP implementation by calculating the exact multipole expansion for a lossless dielectric nanosphere and comparing the results to the Mie theory. The electric field distributions were obtained by FDTD Solutions (Lumerical Inc.). The simulation setup is schematically shown in Figure 2a. The resonator is a lossless nanosphere with a refractive index of 4. A diameter was set to 180 nm to exhibit dipolar and quadrupolar Mie resonances in the visible spectrum. To capture the electric field distributions, a 3D discrete Fourier transform (DFT) monitor (230 nm×230 nm×230 nm) was utilized together with a 3D index monitor that has the same dimension. As a light source, *x*-polarized plane wave was injected along the *z*-axis using a total-field scattered-field (TFSF) source (280 nm×280 nm×280 nm). The simulation domain (1.2 μm×1.2 μm×1.2 μm) were defined by perfectly matched layers in all boundaries. In the domain, Yee cells were automatically defined by a graded mesh, except for a region around the resonator where a 4 nm square mesh was overridden. Additionally, a closed monitor consisting of six DFT monitors were added in the scattered field region to measure total scattering cross sections in simulation. After the simulation, multipole contributions were computed with MENP.

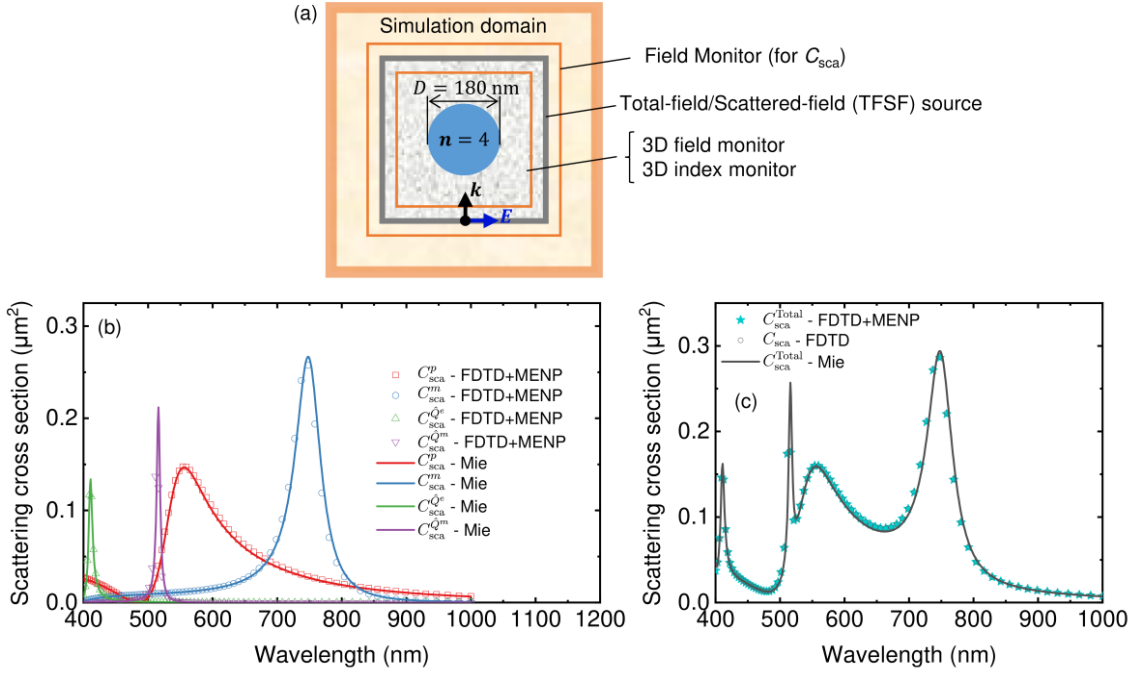

**Figure 2.** Benchmark of exact.m for a lossless nanosphere with a refractive index $n = 4$ and a diameter $D = 180$ nm. (a) Schematic illustration of a simulation setup in the full-field simulation. (b,c) Calculated scattering cross sections (b) from each multipole contribution and (c) a total of them (plots with symbols). For comparison, those calculated with a rigid analytical solution, the Mie theory, is shown with solid lines.

Scattering cross sections obtained with FDTD and MENP (exact.m) are shown in Figure 2b with symbols. For comparison, analytically calculated spectra, the Mie theory, for the same configuration are shown with solid lines. One can see quantitative agreement for all the spectra. The nice agreement can be attributed to the exact expression of the multipole expansion beyond the long-wavelength approximation; the approximated computation results in deviation, especially in the short wavelength range. Note that the trivial deviation, such as a slight blue-shift of a magnetic dipole resonance, can be further reduced by increasing the mesh accuracy in FDTD (*i.e.,* not the error resulted from the computation in MENP). Figure 2c compares total scattering obtained by three approaches. The plots with stars were calculated by summing up partial scattering cross sections calculated by MENP. Circles were the data obtained directly in FDTD using a monitor surrounding the TFSF source. The solid line is the spectrum obtained from the Mie theory into which up to 10th order resonances are included. Likewise, the spectra show good agreement.

We move on to a benchmark for a resonator made of realistic material. Here, we applied `exact.m` to a silicon nanosphere with a diameter of 200 nm. The complex refractive indices were adopted from literature by E. Palik.[40] A simulation setup similar to that for Figure 2 was constructed, but the mesh overriding around the resonator was reduced to 8 nm. The simulated plots (symbols) in Figure 3 perfectly tracks the analytical spectra (solid lines) as well. One can now see that the higher-order dipole resonances around the wavelength range from 400 to 500 nm are adequately reproduced.

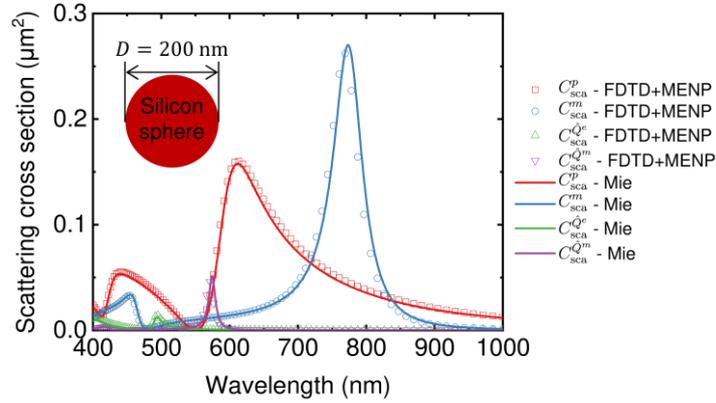

**Figure 3.** Benchmark for scattering cross sections of a silicon nanosphere with a complex refractive index[40] and diameter $D = 200$ nm. The multipole contributions computed with FDTD and MEMP and those obtained with the Mie theory are plotted with colored symbols and solid lines, respectively.

As the implementation has been validated, MENP can now be applied to arbitrary structures. As a demonstration for a specific application in nanophotonics, we demonstrate a multipole analysis on the anapole state. The structure here consists of a silicon nanodisk with a diameter and height of 310 and 50 nm, respectively, as shown in Figure 4a. As described above, the analysis of the anapole state requires the introduction of a toroidal dipole moment. Accordingly, `toroidalME.m` was applied to the electric field distributions computed in the same manner. The computed partial scattering cross sections are plotted in Figure 4b and are in accordance with those in literature.[24] Different from the result based on exact expressions, the total scattering is not just a summation of multipolar contributions and exceeded by the electric dipolar scattering. This is because the electric and the toroidal dipoles have crosstalk as shown in eq. 6, which leads to the constructive interference of the radiation. In this way, the dip in total scattering (black band in Figure 4b) occurs when the contributions of electric and toroidal dipoles match (*i.e.*, anapole state). For the anapole condition, the phases of the moments are also important to cancel out the radiation perfectly. The phase of multipole moments can be extracted as a complex argument in computation. Figure 4c shows the phases of the electric dipole and toroidal one (as $-ik\mathbf{T}$). One can see that the lines cross each other around the anapole wavelength, and thus the anapole condition ($\mathbf{p} = -ik\mathbf{T}$) is satisfied. A supplementary code toroidalME_phase.m is also provided in the MENP package. Although we demonstrated the features of MENP for dielectric Mie resonators, it can also be applied to plasmonic materials.

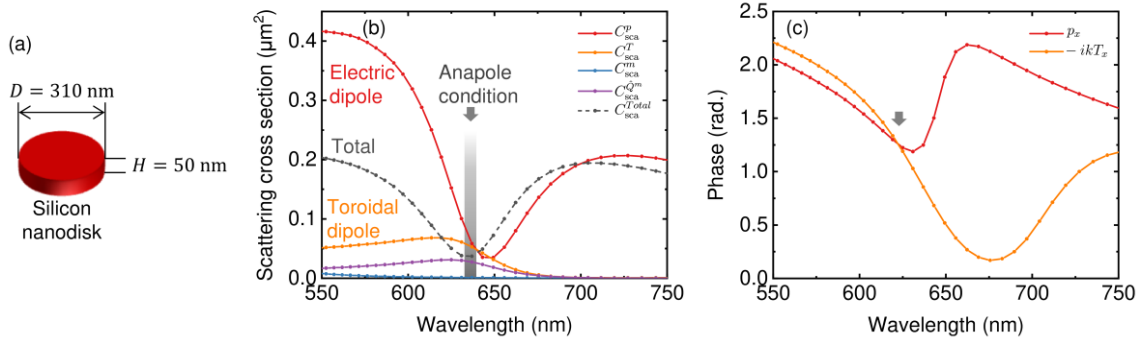

**Figure 4.** Multipole analysis for a silicon nanodisk by toroidalME.m. (a) Schematic of a simulation model composed of a silicon nanodisk with a diameter of 310 nm and a height of 50 nm. (b) Scattering cross sections of total and multipole contributions obtained by FDTD and MENP. The anapole state, a minimum of the total scattering due to the suppression of electric and toroidal dipolar scatterings, is shown by a dark line. (c) Plots of phases, that is the arguments of complex electric ($p_x$, red line) and toroidal dipole ($-ikT_x$, orange line) moments. The anapole condition ($\mathbf{p} = -ik\mathbf{T}$) is satisfied at the wavelength designated by an arrow.

## 4. SUMMARY AND OUTLOOK

In summary, we have presented an open-source MATLAB program MENP designed for efficient computation of multipole expansion in nanophotonics. The program is based on the recently developed exact expression of a multipole expansion and provides the post-analysis for commonly utilized simulation software such as FDTD and FEM. The validity and correctness of the program are demonstrated by comparing the results with analytically obtained quantities. The implementation of multipole expansion under the long-wavelength approximation is also presented to explain the excitation of the anapole state with the help of toroidal dipole moment in a nanodisk. Given the increasing interest in various multipolar interferences in nanophotonics, MENP may have a significant impact on nanophotonics as well as plasmonics community by helping researchers to interpret the physical meaning of scattering-based phenomena.

**ACKNOWLEDGEMENT**

T. H. acknowledges the support under Grant-in-Aid for JSPS Research Fellows. This work was partly supported by JSPS KAKENHI Grants 18J20276 and 18KK0141.

**REFELENCE**

[1] Y. Kivshar, Natl. Sci. Rev. 5 (2018) 144–158.

[2] S. Jahani, Z. Jacob, Nat. Nanotechnol. 11 (2016) 23–36.

[3] A.I. Kuznetsov, A.E. Miroshnichenko, M.L. Brongersma, Y.S. Kivshar, B. Luk'yanchuk, Science 354 (2016) aag2472.

[4] R. Won, Nat. Photonics 13 (2019) 585–587.

[5] I. Staude, T. Pertsch, Y.S. Kivshar, ACS Photonics 6 (2019) 802–814.

[6] W. Liu, Y.S. Kivshar, Philos. Trans. R. Soc. A Math. Phys. Eng. Sci. 375 (2017) 20160317.

[7] A.B. Evlyukhin, S.M. Novikov, U. Zywietz, R.L. Eriksen, C. Reinhardt, S.I. Bozhevolnyi, B.N. Chichkov, Nano Lett. 12 (2012) 3749–3755.

[8] A.I. Kuznetsov, A.E. Miroshnichenko, Y.H. Fu, J. Zhang, B. Luk'yanchuk, Sci. Rep. 2 (2012) 492.

[9] T. Hinamoto, S. Hotta, H. Sugimoto, M. Fujii, Nano Lett. 20 (2020) 7737–7743.

[10] S. Kruk, Y. Kivshar, ACS Photonics 4 (2017) 2638–2649.

[11] M. Khorasaninejad, W.T. Chen, R.C. Devlin, J. Oh, A.Y. Zhu, F. Capasso, Science 352 (2016) 1190–1194.

[12] S. Yi, M. Zhou, Z. Yu, P. Fan, N. Behdad, D. Lin, K.X. Wang, S. Fan, M. Brongersma, Nat. Nanotechnol. 13 (2018) 1143–1147.

[13] P. Fan, U.K. Chettiar, L. Cao, F. Afshinmanesh, N. Engheta, M.L. Brongersma, Nat. Photonics 6 (2012) 380–385.

[14] A.F. Cihan, A.G. Curto, S. Raza, P.G. Kik, M.L. Brongersma, Nat. Photonics 12 (2018) 284–290.

[15] S.T. Ha, Y.H. Fu, N.K. Emani, Z. Pan, R.M. Bakker, R. Paniagua-Domínguez, A.I. Kuznetsov, Nat. Nanotechnol. 13 (2018) 1042–1047.

[16] O. Yavas, M. Svedendahl, R. Quidant, ACS Nano 13 (2019) 4582–4588.

[17] H. Sugimoto, T. Okazaki, M. Fujii, Adv. Opt. Mater. 2000033 (2020) 2000033.

[18] P.R. Wiecha, A. Cuche, A. Arbouet, C. Girard, G. Colas Des Francs, A. Lecestre, G. Larrieu, F. Fournel, V. Larrey, T. Baron, V. Paillard, ACS Photonics 4 (2017) 2036–2046.

[19] Y. Yang, A.E. Miroshnichenko, S. V. Kostinski, M. Odit, P. Kapitanova, M. Qiu, Y.S. Kivshar, Phys. Rev. B 95 (2017) 2–11.

[20] Y.H. Fu, A.I. Kuznetsov, A.E. Miroshnichenko, Y.F. Yu, B. Luk'yanchuk, Nat. Commun. 4 (2013) 1527.

[21] R. Alaee, R. Filter, D. Lehr, F. Lederer, C. Rockstuhl, Opt. Lett. 40 (2015) 2645.

[22] H. Sugimoto, T. Hinamoto, M. Fujii, Adv. Opt. Mater. 1900591 (2019) 1900591.

[23] H.K. Shamkhi, K. V. Baryshnikova, A. Sayanskiy, P. Kapitanova, P.D. Terekhov, P. Belov, A. Karabchevsky, A.B. Evlyukhin, Y. Kivshar, A.S. Shalin, Phys. Rev. Lett. 122 (2019) 193905.

[24] A.E. Miroshnichenko, A.B. Evlyukhin, Y.F. Yu, R.M. Bakker, A. Chipouline, A.I. Kuznetsov, B. Luk'yanchuk, B.N. Chichkov, Y.S. Kivshar, Nat. Commun. 6 (2015) 8069.

[25] B. Luk'yanchuk, R. Paniagua-Domínguez, A.I. Kuznetsov, A.E. Miroshnichenko, Y.S. Kivshar, Philos. Trans. R. Soc. A Math. Phys. Eng. Sci. 375 (2017) 20160069.

[26] J.S. Totero Gongora, G. Favraud, A. Fratalocchi, Nanotechnology 28 (2017) 104001.

[27] K. V. Baryshnikova, D.A. Smirnova, B.S. Luk'yanchuk, Y.S. Kivshar, Adv. Opt. Mater. 7 (2019) 1801350.

[28] S.-Q. Li, K.B. Crozier, Phys. Rev. B 97 (2018) 245423.

[29] G. Grinblat, Y. Li, M.P. Nielsen, R.F. Oulton, S.A. Maier, Nano Lett. 16 (2016) 4635–4640.

[30] L. Xu, M. Rahmani, K. Zangeneh Kamali, A. Lamprianidis, L. Ghirardini, J. Sautter, R. Camacho-Morales, H. Chen, M. Parry, I. Staude, G. Zhang, D. Neshev, A.E. Miroshnichenko, Light Sci. Appl. 7 (2018) 44.

[31] F. Monticone, D. Sounas, A. Krasnok, A. Alù, ACS Photonics 6 (2019) 3108–3114.

[32] L. Wei, Z. Xi, N. Bhattacharya, H.P. Urbach, Optica 3 (2016) 799.

[33] J.A. Parker, H. Sugimoto, B. Coe, D. Eggena, M. Fujii, N.F. Scherer, S.K. Gray, U. Manna, Phys. Rev. Lett. 124 (2020) 097402.

[34] R. Alaee, C. Rockstuhl, I. Fernandez-Corbaton, Adv. Opt. Mater. 7 (2019) 1800783.

[35] R. Alaee, C. Rockstuhl, I. Fernandez-Corbaton, Opt. Commun. 407 (2018) 17–21.

[36] P. Grahn, A. Shevchenko, M. Kaivola, New J. Phys. 14 (2012) 093033.

[37] G. Mie, Ann. Phys. 330 (1908) 377–445.

[38] J.D. Jackson, Classical Electrodynamics, 3rd ed., Wiley, 1998.

[39] E.A. Gurvitz, K.S. Ladutenko, P.A. Dergachev, A.B. Evlyukhin, A.E. Miroshnichenko, A.S. Shalin, Laser Photonics Rev. 13 (2019) 1–13.

[40] Edward D. Palik, Handbook of Optical Constants of Solids, Academic Press, San Diego, USA, 1998.